# Nanoscale Graphene Disk: A Natural Functionally Graded Material

**--** The Thermal Conductivity of Nanoscale Graphene Disk by Molecular Dynamics Simulation

Nuo Yang[1a)*], Shiqian Hu[2*], Dengke Ma[1], Tingyu Lu[2], and Baowen Li[2,3,4,5b)]

[1]School of Energy and Power Engineering, Huazhong University of Science and Technology (HUST), Wuhan 430074, People's Republic of China

[2] Center for Phononics and Thermal Energy Science, School of Physics Science and Engineering, Tongji University, 200092 Shanghai, People's Republic of China

[3]Graphene Research Center, National University of Singapore, Singapore 117546, Singapore

[4]NUS Graduate School for Integrative Sciences and Engineering, National University of Singapore, Singapore 117456, Singapore

[5] Department of Physics and Centre for Computational Science and Engineering, National University of Singapore, Singapore 117542, Singapore

*N. Yang and S. Hu contributed equally to this work.

[a)] Electronic mail: nuo@hust.edu.cn

[b)] Electronic mail: phylibw@nus.edu.sg

**ABSTRACT**

In this letter, we investigate numerically (by non-equilibrium molecular dynamics) and analytically the thermal conductivity of nanoscale graphene disks (NGDs), and discussed the possibility to realize FGM with only one material, NGDs. We found that the NGD has a graded thermal conductivity and can be used as FGM in a large temperature range. Moreover, we show the dependent of NGDs' thermal conductivity on radius and temperature. Our study may inspire experimentalists to develop NGD based FGMs and help heat removal of hot spots on chips by graphene.

**Introduction**

Generally, a functionally graded material (FGM) is a composite, consisting of two or more phases, which is fabricated such that its composition varies in some spatial direction.[1,2] The concept of FGMs was proposed in 1984 by material scientists as a means of preparing thermal barrier materials.[3] This design is intended to take advantage of certain desirable features of each of the constituent phases. For example, FGMs can be used to alleviate the harmful effects of heat on structures. The FGM may usually consist of pure ceramic, at the hotter end, because of the ceramic's higher thermal resistance. In contrast, the cooler end may be pure metal because of its better mechanical and thermal conductivity.

Recently, another attempt to apply the FGM concept to the enhancement of thermoelectric (TE) energy conversion efficiency has been initiated.[4-6] TE materials are important for generating electricity from waste heat and being used as solid-state Peltier coolers. The performance of thermoelectric materials depends on the figure of merit (ZT), $ZT = S^2 \sigma T/\kappa$, where S, T, $\sigma$, and $\kappa$ are the Seebeck coefficient, absolute temperature, electrical conductivity and total thermal conductivity, respectively. A functional graded thermoelectric material (FGTEM) can maintain high values of ZT over a wide range of temperatures, for example, by controlling a gradual change of dopant concentration along the length of a TE device. Muller et al. reviewed various methods for fabrication of FGTEMs and how the gradient impacts the resulting efficiencies.[5]

The graphene is a two-dimensional structure consisting of a single atomic layer of carbon.[7] The thermal properties of graphene have attracted immense interests recently,[8-19] because the size dependence[8,9] and the super-high value[10,11] of thermal conductivity have been observed. Because phonons (lattice vibrations) dominate thermal transport in graphene, it raised the exciting prospect to use graphene-based nano-sheets as thermal (phononics) devices.[20-23] Besides, the management of phonons provide advances in thermal devices, such as thermal diodes,[24] thermal cloaking,[25,26] thermoelectrics,[27-29] and thermocrystals.[30-32]

In this letter, we investigate numerically and analytically the thermal conductivity of nanoscale graphene disks (NGDs), and discussed the possibility to realize FGM with only one material, NGDs. Moreover, we show the dependence of NGDs' thermal conductivity on radius and temperature. Our study may inspire experimentalists to develop NGD based FGMs and help heat removal of hot spots on chips by graphene.

**MD Simulation Methods**

A NGD structure is shown in figure 1, where the difference between $r_{in}$ and $r_{out}$ is defined as L. The lattice constant (a) and thickness (d) of NGD are 0.1418 nm and 0.334 nm, respectively. As we study the thermal conductivity of NGD by using classical non-equilibrium molecular dynamics (NEMD) method, a temperature gradient is building in NGD along the radial direction.

The NGD can be look as a serial rings ($1^{st}$, $2^{nd}$,...,$N^{th}$) whose thickness is defined as a. In order to establish a temperature gradient, the atoms from the $2^{nd}$ to $4^{th}$ rings and the atoms in the $(N-1)^{th}$ ring are controlled by Nosé-Hoover heat baths[20] with temperatures $T_H$ and $T_L$, respectively. In some low dimensional structures, Nosé-Hoover heat baths is not sufficiently chaotic.[33] To ensure our results are independent of heat bath, Langevin heat bathes[34] are also used. The results by both types of heat baths give rise to the graded thermal conductivity (details of results by Langevin in supporting information I). The atoms at boundaries (the 1st and Nth rings) are fixed.

The potential energy is described by a Morse bond and a harmonic cosine angle for bonding interaction, which include both two-body and three-body potential terms.[35,36] Although this force field potential is developed by fitting experimental parameters for graphite, it has been testified by the calculation of thermal conductivity of carbon nanotubes.[37] To integrate the discretized differential equations of motions, the velocity Verlet algorithm is used. The MD simulation time step, $\Delta t$, is chosen as 0.5 fs. Simulations are carried out long enough (2 ns) to guarantee that the system would reach a steady state. Then, the kinetic temperature ($T = \sum_i m_i v_i^2/2$) at each ring and the heat flux in each thermal bath are averaged over 3 ns.

The heat flux (J) transferred across the each ring can be calculated at the heat bath region as

$$J_{T_H(T_L)} = \frac{1}{N_{T_H(T_L)}} \sum_{i=1}^{N_{T_H(T_L)}} \frac{\Delta \varepsilon_i}{2\Delta t} \tag{1}$$

where $\Delta \varepsilon$ is the energy added to/removed from each heat bath ($T_H$ or $T_L$) at each step $\Delta t$. The thermal conductivity ($\kappa$) are calculated based on the Fourier definition as

$$J = -2\pi \kappa r \mathrm{d} \frac{dT}{dr} \tag{2}$$

where r is the radius of each ring. We use a combination of time and ensemble sampling to obtain better average statistics. The results represent averages from 12 independent simulations with different initial conditions. Each case runs longer than 3 ns after the system reached the steady state.

**MD and analytical results**

For a uniform disk of bulk material, the thermal conductivity ($\kappa$) is defined as

$$\kappa = J \ln(r_{out}/r_{in}) / 2\pi d (T_{in} - T_{out}). \tag{3}$$

In the calculation of $\kappa$, the temperature gradient, dT/dr, is an important factor. As shown in figure 2(a), different from bulk disk structures, the temperature gradient of NGDs is not a constant and depends on radius r. The values for different r are obtained directly from the discrete temperature profiles of MD. Then, we calculated the NGD's thermal conductivity with different outer radius at room temperature shown in figure 2(b). It is obviously that the values of thermal conductivity of each NGD are, instead of constants, dependent on the radius. That is, the NGD is a special structure with a graded thermal conductivity.

As the curves of $\kappa$ on a log-log plot are linear, we can write:

$$\kappa(r) = \kappa_0 \left[ \frac{\ln(C/r)}{\ln(C/r_{out})} \right]^\alpha \tag{4}$$

where $\alpha$ would depend on temperature and geometric size of NGDs, $\kappa_0$ and C are constants.

Similar to the solutions in a hollow circular cylinder FGM,[2] we derive the analytical results of non-homogeneous steady state temperature distributions in NGDs (the concrete derivation processes are given in the Supporting information II). The solutions are as:

$$T(r) = \begin{cases} T(r_{in}) + [T(r_{out}) - T(r_{in})] \dfrac{[\ln(r/C)]^{1-\alpha} - [\ln(r_{in}/C)]^{1-\alpha}}{[\ln(r_{out}/C)]^{1-\alpha} - [\ln(r_{in}/C)]^{1-\alpha}}, & \alpha \neq 1 \\ \\ T(r_{in}) + [T(r_{out}) - T(r_{in})] \dfrac{\ln[\ln(r/C)] - \ln[\ln(r_{in}/C)]}{\ln[\ln(r_{out}/C)] - \ln[\ln(r_{in}/C)]}, & \alpha = 1 \end{cases} \tag{5}$$

We choose the value of C as $r_{in}/e$ for the purpose of normalization in figures.

Then, we investigate the temperature and radius effect on the thermal conductivity of NGD by MD simulations. Firstly, we choose a disk with the outer radius as 13.89 nm and calculate the thermal conductivity of different temperatures from 600 K to 2500 K. The temperature profiles along the radial direction are plotted in figure 3(a). It is shown that the numerical data can be well fitted by our analytical results in Eq. (5), which give evidence of our assumption of graded thermal conductivity in Eq. (4). The fitted temperature gradient has a much smaller fluctuation comparing to the data extracted directed from the discrete temperature profiles of MD. The fitted values of $\alpha$ decreased toward zero as the temperature increasing. That is, for the high temperature limit, the temperature profile in NGD would be close to linear and there would be no graded thermal conductivity.

As shown in figure 3(b), the thermal conductivities of NGD are calculated with different temperatures from 300 K to 2500 K. When the temperature is 300 K, the value of thermal conductivity of the outermost ring is nearly four times larger than the value of innermost ring. As the temperature increases to 2500 K, the value of $\kappa(r_{out})$ is almost as twice as that of $\kappa(r_{in})$. It is clear shown that the NGD structures have graded thermal conductivity and can be used as FGM in a large temperature range, which is the main result of this paper.

Besides the temperature effect, we also investigate the size effect on thermal conductivity of NGD at 1000 K, where the $r_{out}$ are changed from 8.22 to 25.24 nm. The temperature profiles along the radial direction with different outer radius are shown in figure 4(a), where the temperature profile by MD can be well fitted by our analytical results in Eq.

(5). The thermal conductivity of NGDs with the different $r_{out}$ are calculated and plotted in figure 4(b). The results shows that the values of $\alpha$ is not sensitive of $r_{out}$. However, the values $\kappa_0$ depends on the $r_{out}$. That is, the value of thermal conductivity is enhanced with the increase of NGD's outer radius.

**Discussions and conclusion**

Based on our MD simulation results of the NGD's thermal conductivity ($\kappa$), we found that $\kappa(r)$ depends on radius and increase gradedly with r as shown in Eq. (4),

$$\kappa(r) = \kappa_0 \left[ \frac{\ln(C/r)}{\ln(C/r_{out})} \right]^{\alpha}.$$ The value of $\alpha$ would depend on temperature and goes to zero at high temperature limit. That is, there would be no graded effect in NGDs in the high temperature limit where the nonlinear effects are dominant and there are much more phonon-phonon couplings. The value of $\kappa_0$ depends the geometric size of NGDs when temperature is invariable. The larger is the outer radius of NGD, the higher is the value of $\kappa_0$. The trend is accordant with the phonon transport in nanostructures,[20] where there are larger mean free path and more eigenmodes in a larger structure.

Recently, the size and confinement effect in thermal properties of nanostructures are well understood.[20,29,38-40] The thermal conductivity of nanostructures depends on the size[41-44] and the Fourier' law is not valued in nanoscale. A nanoscale graphene disk can be looked as a serial of thin graphene rings with different radius from $r_{in}$ to $r_{out}$. We calculated the power spectra of several rings with different radius in figure (5), to understand the underlying mechanism in nanoscale. The spectra are obtained by Fourier transforming velocity autocorrelation function. Due to the confinement of size, the phonon wavelength in a ring is limited from the lattice constant to the circumference, which leads to the confinement in phonon's modes shown in the spectra of nanoscale rings obviously. That is, the number of peaks, which corresponding to egienmodes, is much decreased as the radius lessening. Analogy to nanowires and nanotubes with different length,[41-44] the ring with different radius will have different ability in heat transfer. For a smaller ring, the thermal conductivity will be lower due to the very small number of egienmodes. With the increase of radius, more and more phonons, especially phonons with longer wavelengths, are existed in the ring, which will result in the increase of thermal conductivity. When the rings are connected one by one, the structure is nonhomogenous along the direction of thermal gradient, and a graded structure will be build.

In summary, by using the classical nonequilibrium MD method, we have investigated the radius and temperature effect on the thermal conductivity of NGD. It is found that the NGD is a natural structure with a graded thermal conductivity, without any artificial compounding. The thermal property of NGDs can be modulated by changing the temperature and the size. Our analytical results of the temperature profile in NGD, Eq. (5), can fit simulation results very well. It is shown that our prediction of Eq. (4) is reasonable. Our study may inspire experimentalists to develop NGD based FGMs and help heat removal of hot spots on chips or enhancement of thermoelectric energy conversion efficiency by graphene.

ASSOCIATED CONTENT

**Supporting Information**

Simulation and calculation details. This material is available free of charge via the Internet.

**Acknowledgements**

B.L. was supported in part by the grants from MOE Grant R-144-000-305-112 of Singapore, and the National Natural Science Foundation of China (11334007). N.Y. was sponsored in part by the grants from Self-Innovation Foundation (2014TS115) of HUST, Talent Introduction Foundation (0124120053) of HUST, and the National Natural Science Foundation of China Grant (11204216). The authors are grateful to Nianbei Li, Jun Zhou, Xiangfan Xu, and Yunyun Li for useful discussions. The authors thank the National Supercomputing Center in Tianjin (NSCC-TJ) for providing help in computations.

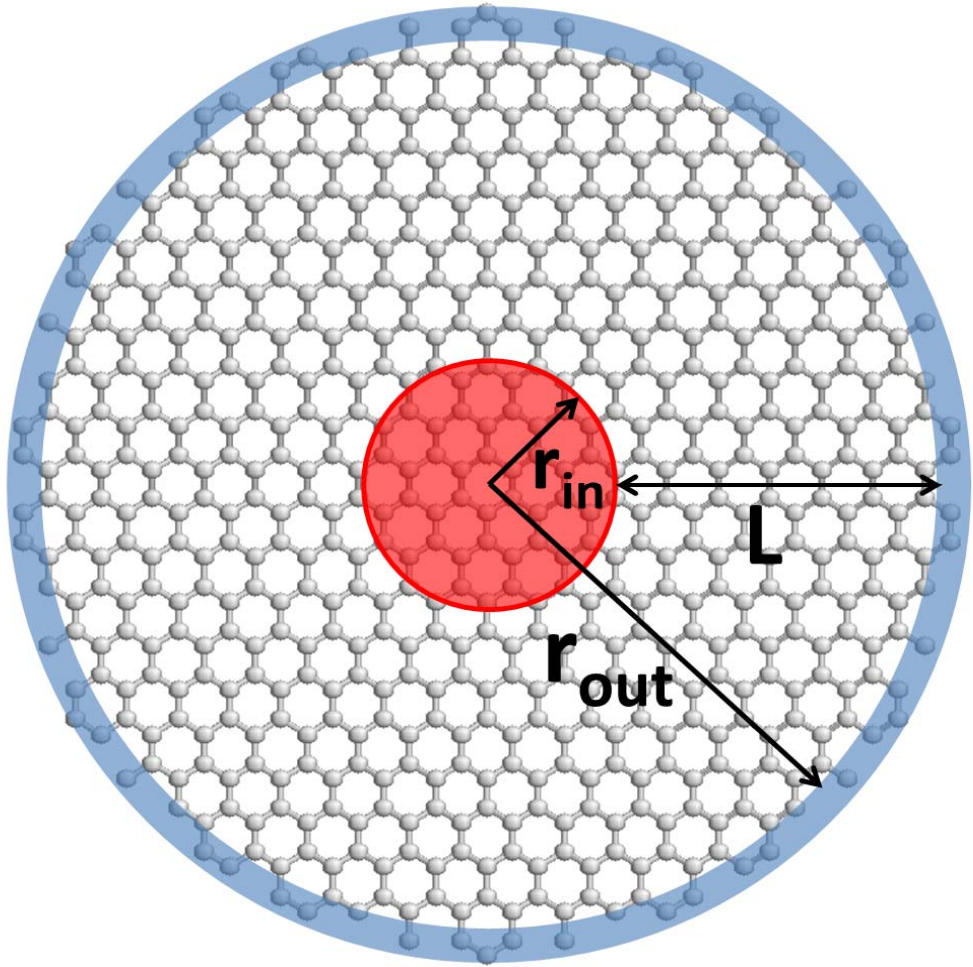

Figure 1. Schematic pictures of the graphene disk structures. The atoms from the 2nd to 4th rings and the atoms in the (N-1)th ring are controlled by Nosé-Hoover heat baths with temperatures $T_H$ and $T_L$, respectively. The atoms at boundaries (the 1st and Nth rings) are fixed.

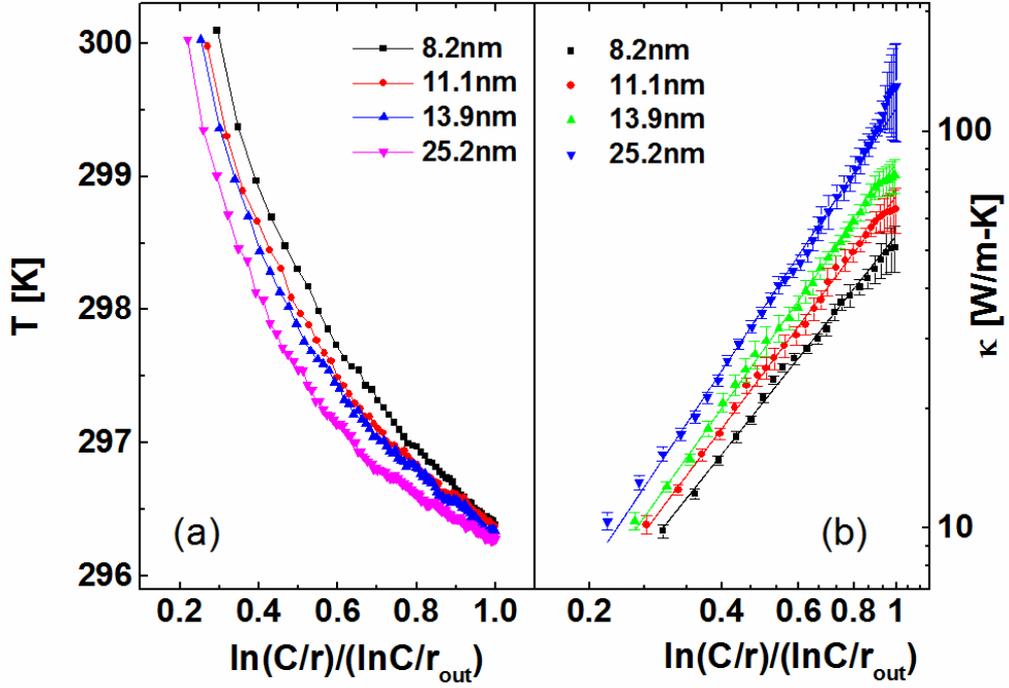

Figure 2. (Color on-line) (a) Temperature profiles in the radial direction with different outer radius ($r_{out}$) at 300 K. (b) The thermal conductivity of graphene disk with different outer radius ($r_{out}$) at 300 K. The symbols are numerical data and the lines are fitted lines. The fitted values $\alpha$ are 1.38±0.03, 1.47±0.01, 1.55±0.01, 1.66±0.02 corresponding to rout as 8.22, 11.06, 13.89, and 25.24 nm, respectively. The error bar is standard deviation of 12 MD simulations with different initial conditions.

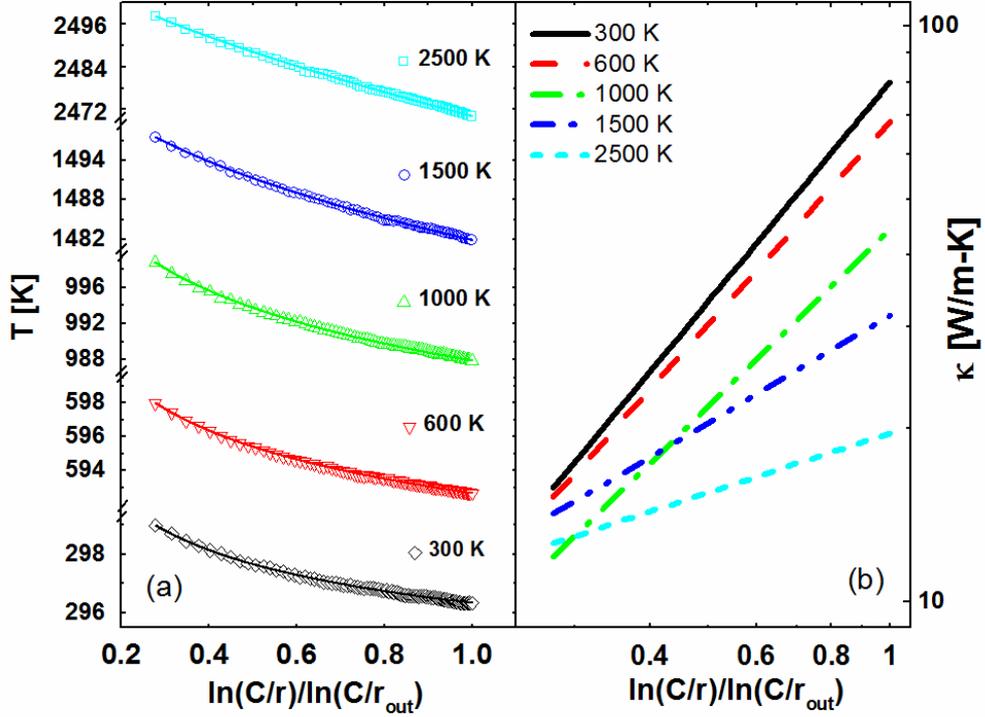

Figure 3. (Color on-line) We choose the value of C as $r_{in}/e$ for the purpose of normalization in horizontal ordinate. (a) The temperature profiles of nanoscale graphene disk (NGD) along the radial direction with different temperatures. The outer radius of NGD equals 13.89 nm. The symbols are MD simulation results and the fitted red lines are based on our analytical results Eq. (5). The fitting value $\alpha$ is $1.26\pm0.01$, $1.17\pm0.01$, $1.02\pm0.01$, $0.62\pm0.01$, $0.34\pm0.01$ corresponding to the temperature is 300, 600, 1000, 1500, 2500 K, respectively. (b) The thermal conductivity of graphene disk with temperature from 600 K to 2500 K . The graded thermal conductivities have form as shown in Eq. (3). .

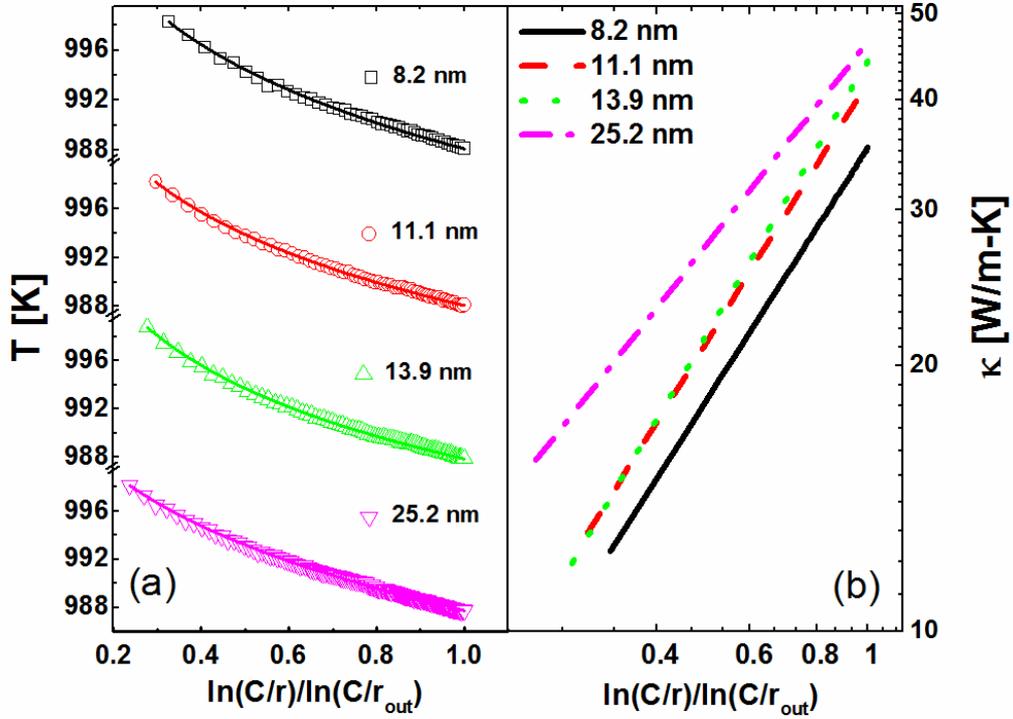

Figure 4. (Color on-line) We choose the value of C as $r_{in}/e$ for the purpose of normalization in horizontal ordinate. (a) The temperature profiles in the radial direction with different outer radius ( $r_{out}$ from 8.22 to 25.24 nm ) at 1000 K. The symbols are numerical data and the lines are analytical data. The value $\alpha$ is 0.94±0.02, 0.97±0.01, 1.02±0.01, 0.76±0.01 corresponding to $r_{out}$ 8.22, 11.06, 13.89, 25.24 nm, respectively (b) The thermal conductivity of graphene disk with different outer radius ( $r_{out}$ from 8.22 to 25.24 nm) at 1000 K, which is plotted on log-log scale.

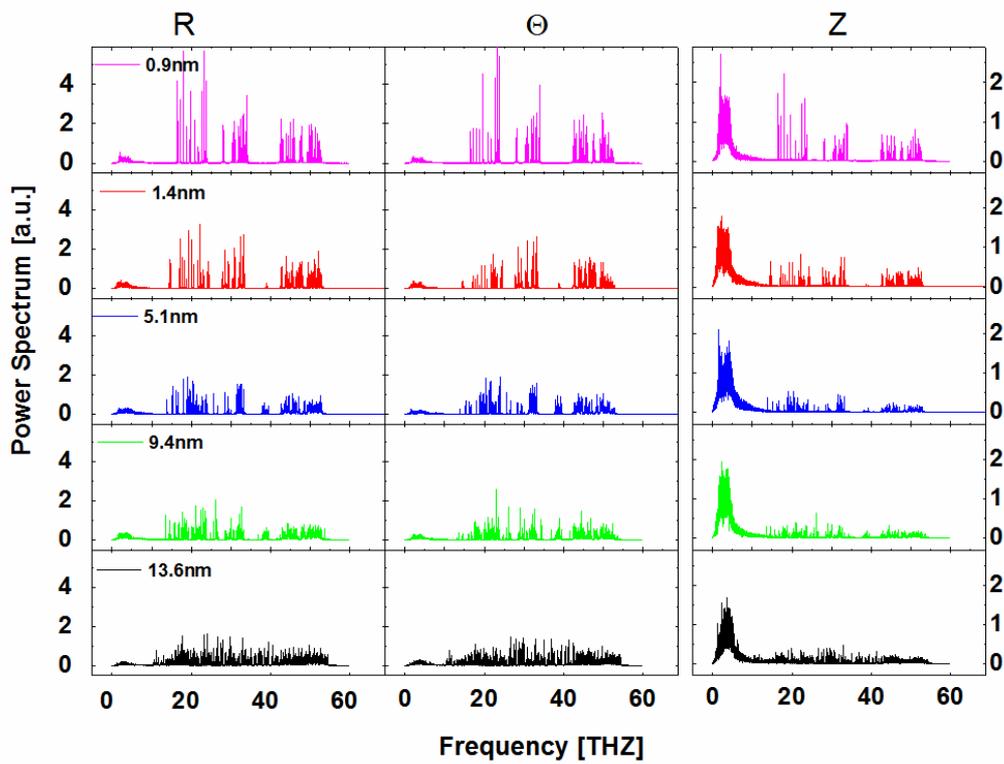

Figure 5. (Color on-line) Normalized power spectra of graphene rings along two in-plane directions (r and θ) and one out-plane direction (z). The radiuses of rings are 0.9 nm, 1.4 nm, 5.1 nm, 9.4 nm, and 13.7 nm.